\documentclass[aps,prb,reprint,superscriptaddress,floatfix,nobibnotes]{revtex4-1}

\usepackage{graphicx}
\usepackage{geometry}
\usepackage{amssymb,amsfonts,amsmath,textcomp}
\usepackage{color}
\usepackage{pbox}
\usepackage{hyperref}

\newcommand{\Vect}[1]{\mathbf{#1}}
\newcommand{\LT}[1]{\tilde{#1}}
\newcommand{\FT}[1]{\hat{#1}}
\newcommand{\InertialMW}{shear inertia model }
\newcommand{\stdvisc}{\bar{\eta}}
\renewcommand{\imath}{i}

\begin{document}

\title{Molecular interpretation of the non-Newtonian viscoelastic behavior of liquid water at high frequencies}

\author{Julius C. F. Schulz}
\affiliation{Freie Universit\"at Berlin, Fachbereich Physik, 14195 Berlin, Germany}
\author{Alexander Schlaich}
\affiliation{Freie Universit\"at Berlin, Fachbereich Physik, 14195 Berlin, Germany}
\affiliation{Universit\'e Grenoble Alpes, CNRS, LIPhy, 38000 Grenoble, France}
\author{Matthias Heyden}
\affiliation{School of Molecular Sciences and Center for Biological Physics, 
Arizona State University, Tempe, Arizona 85287-1604, USA}
\author{Roland R. Netz}
\affiliation{Freie Universit\"at Berlin, Fachbereich Physik, 14195 Berlin, Germany}
\author{Julian Kappler}
\email[]{jkappler@physik.fu-berlin.de}
\affiliation{Freie Universit\"at Berlin, Fachbereich Physik, 14195 Berlin, Germany}
\affiliation{Department of Applied Mathematics and Theoretical Physics, University of Cambridge, Cambridge CB3 0WA, UK}

\date{\today}

\begin{abstract}
Using classical as well as ab-initio molecular dynamics simulations,
we calculate the frequency-dependent shear viscosity of
pure water and water-glycerol mixtures.
In agreement with recent experiments, we find deviations from 
\textcolor{black}{Newtonian-fluid behavior} in the THz regime.
Based on an extension of the  Maxwell model, 
\textcolor{black}{we 
introduce a viscoelastic model} to describe  the observed
viscosity spectrum of pure water.
We find four relaxation modes in the spectrum which we attribute to i) hydrogen-bond network topology
changes, ii) hydrogen-bond stretch vibrations of water pairs,
 iii) collective vibrations of water molecule triplets, and iv) librational excitations
of individual water molecules.
Our model quantitatively describes
the viscoelastic response of liquid water on short timescales, where
the hydrodynamic description via a Newtonian-fluid model breaks down.
\end{abstract}


\maketitle

\section{Introduction}\label{introduction}

Liquid water is an ubiquitous medium on earth, and of fundamental
importance for all organisms\cite{2486906/W6K5BS7C,2486906/TSXWAXKJ}.
A standard model for the large-scale hydrodynamics of liquid water
is the Newtonian fluid\cite{2486906/B3VJUC58,2486906/NKA76ANX,2486906/ZGJTB2CT},
where
one assumes a linear relationship between local instantaneous stresses
and rates of strain, with the
 proportionality constant given by the viscosity.
The combination of the momentum conservation equation
 with this relation
is then known as the Navier-Stokes equation,
and it is the basis for hydrodynamics.

Despite its success for describing dynamics of water and other liquids, this
model has a limited range of applicability.
At high frequencies, when time scales are comparable to those of molecular kinetics
 within the liquid, real water deviates from the Newtonian fluid model.
Slie et al.\cite{2418347/VR5ET3DV} showed by ultrasound absorption measurements that in aqueous
glycerol solutions, at high frequencies the shear viscosity decreases and the
 mixture starts to have an elastic response under shear deformation.
More recently, Pelton et al.\cite{2418347/8BRIQ6QX} showed that the same
occurs for pure water.
Both studies replaced the Newtonian fluid model by a viscoelastic Maxwell fluid\cite{2486906/5SQ8ZI6Z},
which in the low frequency limit reduces to a Newtonian fluid but
can account for the experimentally observed elastic behavior
at high frequencies.
Molecular dynamics simulations of 
 water-glycerol mixtures\cite{lacevic_viscoelasticity_2016} and 
pure water\cite{2486906/M8ANX9ZV,osullivan_viscoelasticity_2019}
also find a non-Newtonian regime at high frequencies,
the onset of which 
is 
 well-described by a Maxwell model\cite{lacevic_viscoelasticity_2016,osullivan_viscoelasticity_2019}.

Understanding this non-Newtonian behavior is becoming increasingly important.
First, with the advancement of nanotechnology, small structures, when in
an aqueous environment, start to probe the regime where the Newtonian-fluid
model of water breaks down\cite{2418347/8BRIQ6QX,ruijgrok_damping_2012}.
Second, the THz regime probed by modern spectroscopic methods
 constitutes the boundary between collective
and single-water dynamics\cite{cunsolo_experimental_1999,ruijgrok_damping_2012,
2486906/MD356JZF,2486906/QU86HKME,2486906/ZW2ASP7U}.
If one seeks detailed insight into dynamics of molecular or collective  processes
of solutes on such fast
time scales, an important
aspect is therefore understanding how water itself behaves on the relevant time and
length scales.

In the present work, we use both
force field molecular dynamics (MD) and ab-initio molecular dynamics (aiMD)
 simulations to probe
the high-frequency behavior of both pure water and water-glycerol mixtures.
From our simulations, we extract the frequency-dependent shear viscoelasticity.
We verify our method by comparing to the experimental results of
Slie et al.\cite{2418347/VR5ET3DV} for the viscoelasticity of water-glycerol mixtures.
Then we investigate the MD spectrum of pure water in more detail.
We propose a viscoelastic model to 
account for deviations of the observed water viscosity spectrum from the Maxwell model.
We identify four independent relaxation modes in the THz regime and link them to molecular 
processes, namely i) hydrogen-bond network topology changes, ii) hydrogen-bond stretch vibrations, 
iii) collective vibrations of water molecule triplets, iv) librational excitations of individual water molecules. 
Our viscosity spectrum based on ab-initio MD (aiMD) simulations
 shows the same high-frequency
features as the spectrum obtained from force field MD simulations,
and thus validates the latter. 
The aiMD spectrum additionally contains features
originating from intramolecular degrees of freedom (OH stretching, OH bending)
at large frequencies
 not included in the
rigid water model used for the force field MD simulations.

\section{Frequency-dependent viscosities}

In this section we recall some
generalizations of the standard Green-Kubo relation
for the shear viscosity\cite{2486906/MCFNDC9T,
2486906/IHJ94W35,
2486906/P3II85DD,
2486906/6I65B8BX,
2486906/454CG5E9,
2486906/9SPF35M5}. 
These generalizations can be used to
calculate the frequency- and wave number dependent shear viscosity in
terms of velocity and stress tensor correlation functions.
Detailed derivations can be found in the Supplemental Material\cite{supplement} (SM) as well
as in the literature\cite{2486906/IHJ94W35,
2486906/P3II85DD,
2486906/6I65B8BX}.

We start from the linearized continuum-mechanical momentum conservation equation\cite{2486906/5SQ8ZI6Z},
\begin{equation}
    \label{eq:LinearizedMomentumConservation}
    \rho \dot{v}_\alpha(\Vect{x},t) = \sum_{\beta=1}^3\partial_\beta \sigma_{\alpha \beta}(\Vect{x},t) ,\qquad \alpha \in \{x,y,z\},
\end{equation}
where $\rho$ is the constant equilibrium volume mass density of the fluid, $\Vect{v}$ its velocity field, $\Vect{\sigma}$ its stress tensor, and a dot denotes the time derivative.
Note that while for a compressible fluid the density is not constant, 
deviations from the equilibrium volume mass density $\rho$
on the left-hand side of Eq.~\eqref{eq:LinearizedMomentumConservation}  would constitute nonlinear
effects and are therefore not considered in our linear treatment.
For a linear, homogeneous, isotropic compressible fluid the stress tensor is given as
\begin{widetext}
\begin{align}
    \label{eq:ConvolutedStressStrain}
    \sigma_{\alpha \beta}(\Vect{x},t) 
    &= 
    - \delta_{\alpha \beta} P(\Vect{x},t) 
    + 2 \int  \int  \eta(|\Vect{x}'|,t')\dot{\epsilon}_{\alpha \beta} (\Vect{x}-\Vect{x}',t-t') 
	  ~ \mathrm{d}^3 \Vect{x}'\, \mathrm{d}t' \\
	& \qquad +  \delta_{\alpha \beta} \sum_{\gamma=1}^3
	 \int  \int \left(\eta'(|\Vect{x}'|,t') - \frac{2}{3}\eta(|\Vect{x}'|,t')\right) \dot{\epsilon}_{\gamma \gamma} (\Vect{x}-\Vect{x}',t-t') 
	  ~ \mathrm{d}^3 \Vect{x}'\, \mathrm{d}t' ,
	  \nonumber
\end{align}
\end{widetext}
where $P$ is the pressure,
$\delta_{\alpha \beta}$ is the Kronecker delta,
 the components of the rate of strain tensor $\dot{\epsilon}$ are
\begin{equation}
    \label{eq:RateOtStrainTensor}
    \dot{\epsilon}_{\alpha \beta} = \frac{1}{2}\left(\frac{\partial v_\alpha}{\partial x_\beta} + \frac{\partial v_\beta}{\partial x_\alpha}\right), \qquad \alpha, \beta \in \{x,y,z\}.
\end{equation}
and $\eta$, $\eta'$ are the shear and volume viscosity kernels, which for an isotropic medium only depend on 
the modulus of the vector $\Vect{x}'$.

If the viscosity kernels decay on length- and time scales that are small compared to 
those on which the rate of strain tensor varies,
the stress tensor defined by Eq.~\eqref{eq:ConvolutedStressStrain} can be approximated as\cite{supplement}
\begin{align}
    \label{eq:StandardStressStrain}
    \sigma_{\alpha \beta}(\Vect{x},t) &\approx 
    - \delta_{\alpha \beta} P (\Vect{x},t)
    + 2 \stdvisc
	  \dot{\epsilon}_{\alpha \beta} (\Vect{x},t) \\
	  \nonumber
	  &\qquad
	 +  \delta_{\alpha \beta} 
	 \left(\stdvisc' - \frac{2}{3} \stdvisc\right) 
	 \sum_{\gamma=1}^3\dot{\epsilon}_{\gamma \gamma} (\Vect{x},t),
\end{align}
where
\begin{align}
	\label{eq:LocalApproximationOfViscosityKernelA}
	\stdvisc &=  \int \int \eta(|\Vect{x}'|,t') \, \mathrm{d}^3 \Vect{x}'\, \mathrm{d}t',\\
	\label{eq:LocalApproximationOfViscosityKernelB}
	\stdvisc' &=  \int \int \eta'(|\Vect{x}'|,t') \, \mathrm{d}^3 \Vect{x}'\, \mathrm{d}t',
\end{align}
are the standard shear and volume viscosities, which do not depend on space and time.
A fluid with stresses given by Eq.~\eqref{eq:StandardStressStrain} is called a Newtonian fluid.
If the stress 
tensor Eq.~\eqref{eq:StandardStressStrain} 
is used in the momentum conservation equation \eqref{eq:LinearizedMomentumConservation}, 
 the linearized compressible Navier-Stokes equation is recovered.
While in this work we mostly consider the spatial average of the shear viscosity kernel,
we are precisely interested in the dynamics on time scales where the approximation
Eq.~\eqref{eq:StandardStressStrain} breaks down,
 and non-Markovian effects become relevant.

From momentum conservation Eqs.~\eqref{eq:LinearizedMomentumConservation}, \eqref{eq:ConvolutedStressStrain}, and the equipartition theorem it follows that the shear viscosity kernel $\eta$ is given
in terms of the trace free part of the stress tensor, 
\begin{equation}
\Pi_{\alpha\beta} =\sigma_{\alpha\beta} - \delta_{\alpha\beta}\,\frac{1}{3}\, \sum_\gamma \sigma_{\gamma \gamma}  \quad \alpha,\beta \in\{x,y,z\},
\end{equation}
as\cite{2486906/P3II85DD,
2486906/454CG5E9,
2486906/9SPF35M5, 
supplement}
\begin{align}
\tilde{\eta}(\Vect{k} = 0, \omega) &= \frac{\beta V}{10} \int_0^{\infty}e^{-\imath\,\omega\,t}\, \sum_{\alpha\beta}\Big\langle  \Pi_{\alpha\beta}(t)\,\Pi_{\alpha\beta}(0) \Big\rangle\,\textrm{d}t,
\label{eq:visc_freq_all}
\end{align}
where $V$ is the volume of the fluid, 
$\beta^{-1} = k_{\mathrm{B}}T$ is the thermal energy 
with $k_{\mathrm{B}}$ the Boltzmann constant and $T$ the absolute temperature, 
the tilde denotes a combined spatial Fourier transform (with wave vector $\Vect{k}$) and temporal half-sided
Fourier transform transform (with angular frequency $\omega$), 
and the average on the right-hand side is to be 
understood as an ensemble average over space and time.
The real part of $\tilde{\eta}$ yields the viscous response under shear, whereas the
imaginary part models the elastic response under shear\cite{2486906/5SQ8ZI6Z}.

Note that, in view of Eqs.~\eqref{eq:StandardStressStrain}, \eqref{eq:LocalApproximationOfViscosityKernelA},
 $\tilde{\eta}$ evaluated at $\Vect{k}=0$ can be thought of as the viscosity kernel for a viscosity that decays on a length scale
 much smaller than the length scale on which $\dot{\epsilon}$ varies, 
but  including memory effects in time. 
If additionally the limit $\omega \rightarrow 0$ is taken, 
memory effects
 in time are also neglected and the Green-Kubo 
relation for $\stdvisc$
 is obtained from Eq.~\eqref{eq:visc_freq_all} as\cite{2486906/MCFNDC9T,
2486906/IHJ94W35,
2486906/P3II85DD,
2486906/454CG5E9,
2486906/6I65B8BX,
2486906/9SPF35M5,
bartosz_memory_2019}
\begin{equation}
\stdvisc = \frac{\beta V}{10} \int_0^{\infty}\, \sum_{\alpha\beta}\Big\langle  \Pi_{\alpha\beta}(t)\,\Pi_{\alpha\beta}(0) \Big\rangle\,\textrm{d}t.
\label{eq:gk_all}
\end{equation}

While Eqs.~\eqref{eq:visc_freq_all}, \eqref{eq:gk_all}, can be used to calculate the viscosity 
within force field MD simulations, where the space-averaged pressure tensor is available\cite{2418347/C7K6AC72},
this is not the case for the ab-initio simulations we also consider in this work.
To calculate shear viscosities also from ab-initio simulations, we use that
for small wave number $k = |\Vect{k}|$, the
 shear viscosity is given by\cite{2486906/IHJ94W35,
2486906/P3II85DD,
2486906/6I65B8BX,
supplement}
\begin{equation}
    \label{eq:ViscosityApproximateFormula2}
    \LT{\eta}({k}, \omega) = -\frac{\rho}{k^2}\int_0^{\infty} e^{-\imath \omega t}~\ddot{\FT{C}}^{\perp}({k},t)~\mathrm{d}t,
\end{equation}
where a hat denotes a spatial Fourier transform and $\FT{C}^{\perp}$ is the 
normalized autocorrelation function of the $\alpha$-component of the transversal velocity,
\begin{align}
	\label{eq:DefinitionTransversalVelocityAutoccorelation}
	\FT{C}^{\perp}({k},t) &= \frac{\langle \FT{v}_{\alpha}^{\perp}(\Vect{k},t) ~\FT{v}_{\alpha}^{\perp}(-\Vect{k},0) \rangle}{\langle \FT{v}_{\alpha}^{\perp}(\Vect{k},0) ~\FT{v}_{\alpha}^{\perp}(-\Vect{k},0) \rangle}, \qquad \alpha \in \{x,y,z\},
\end{align}
with
 $\FT{v}_{\alpha}^{\perp} := \sum_\beta \left(\delta_{\alpha \beta}
-k_{\alpha}k_{\beta}/k^2\right) \FT{v}_{\beta}$  the transversal velocity.
Note that for an isotropic medium,
the right-hand side of Eq.~\eqref{eq:DefinitionTransversalVelocityAutoccorelation}
does not depend on  the component  $\alpha$ used
and only depends on $\Vect{k}$ via the modulus $k$, which is why
$\FT{C}^{\perp}$ has no index $\alpha$ and is written as a function of $k$.
  
In the context of Eq.~\eqref{eq:ViscosityApproximateFormula2}, small wave number means\cite{supplement}
\begin{equation}
    \left|\frac{k^2 \LT{\eta}}{\rho \omega}\right| \ll 1,
    \label{eq:ApproximationCondition}
\end{equation}
which for realistic values for water,
 $|\tilde{\eta}| \approx 0.1~\mathrm{mPa\cdot s}$, $\rho = 10^{3}~\mathrm{kg/m^3}$, 
gives
	$|k|^2 \mathrm{nm}^2/\mathrm{ps} \ll |\omega|$.
Thus, for $|k| = 4~\mathrm{nm}^{-1}$, corresponding to the smallest wave number resolvable in a typical box
in an MD simulation, formula \eqref{eq:ViscosityApproximateFormula2} is valid for
	$\omega \gg 1.6~\mathrm{ps}^{-1}$,
i.e. for frequencies well above 1 THz, which severly limits the applicability of Eq.~\eqref{eq:ViscosityApproximateFormula2}.
While in the limit $\Vect{k} \rightarrow 0$, condition \eqref{eq:ApproximationCondition} is 
always fulfilled and the approximate Eq.~\eqref{eq:ViscosityApproximateFormula2} is valid, 
obtaining $\hat{v}^{\perp}_{\alpha}(\Vect{k}=0,t)$ from simulations is typically difficult because
large simulation boxes are needed to resolve small wave numbers,
and thereby to extrapolate to $\Vect{k} = 0$.

\section{Glycerol Spectra}\label{glycerol-spectra}

\begin{figure}
            \begin{center}
            \includegraphics[width=\columnwidth]{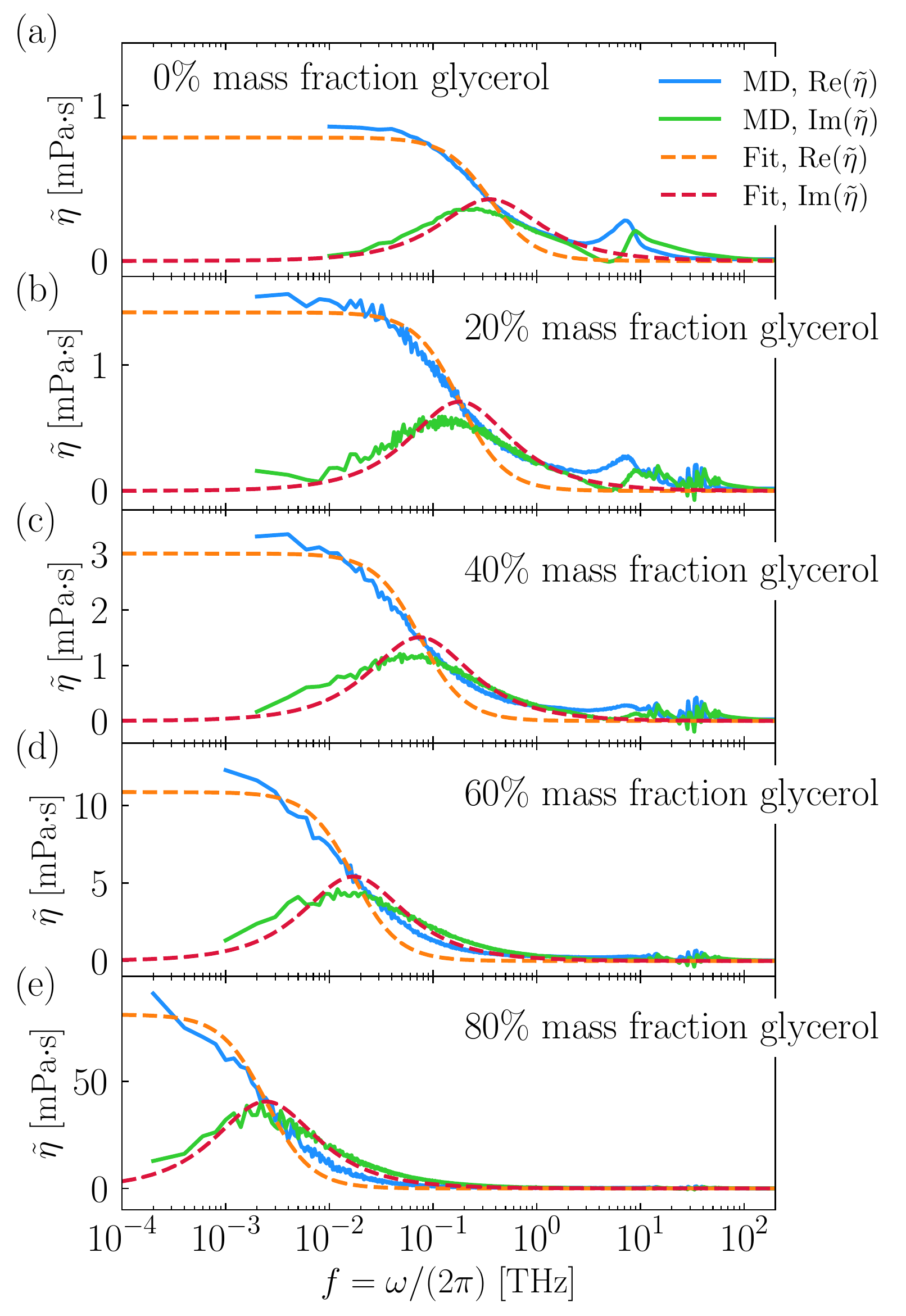}
            \end{center}
            \caption{\textbf{Viscosity spectra $\LT{\eta}$ as calculated from MD simulations of TIP4P/2005
            water at different glycerol concentrations.}
            The real and imaginary parts of the spectra (blue and green solid lines) are calculated from
             the MD data at $\Vect{k} = 0$ using Eq.~\eqref{eq:visc_freq_all}.
            The result of a Maxwell model fit, c.f.~Eq.~\eqref{eq:maxwell},
             to frequencies $10^{-4}\,\mathrm{THz\,}< f < 1\,\mathrm{THz}$,
            is shown as orange and red dashed lines.}
            \label{fig:glyc_spectra}
        \end{figure}

We simulate glycerol solutions\cite{2418347/DTJ5W6BW} with glycerol mass
 fractions 0, 0.2, 0.4, 0.6, 0.8 in TIP4P/2005 water, for details see SM\cite{supplement}.
From the simulations we calculate the 
 respective viscosity spectra at $\Vect{k} = 0$ using Eq.~\eqref{eq:visc_freq_all}.
The resulting spectra are shown as blue and green solid curves in Fig.~\ref{fig:glyc_spectra}.

For low frequencies, the real parts (blue) of the spectra are constant and 
the imaginary parts (green) vanish.
As the frequency is increased, the real parts of the spectra decrease to zero, while
the imaginary parts show peaks.
For low glycerol mass fractions, the real part
of the shear viscosity
 shows a non-monotonic behavior with a second
peak at around 10 THz, accompanied by a peak in the imaginary part at slightly higher frequencies.
For large glycerol mass fractions, this high-frequency peak disappears, and the decay in
 the real part with its corresponding peak in the imaginary
part shift to lower frequencies.

These spectra illustrate the limits of the Newtonian fluid model, as defined by 
Eq.~\eqref{eq:StandardStressStrain}, because for such a fluid one would expect a
 spectrum with constant real part and vanishing imaginary part over the whole frequency range.
Our spectra exhibit deviations from this behavior for high frequencies, indicating that the assumption
of temporal locality breaks down once the rate of strain tensor varies
on the picosecond time scale.

To go beyond Newtonian hydrodynamics, we fit Maxwell models, defined by 
\begin{equation}
 \label{eq:maxwell}
\LT{\eta}(\omega) = \frac{\eta_0}{1-\imath \omega \tau},
\end{equation}
with $\eta_0=\LT{\eta}(0)$ the steady state shear viscosity and a
 timescale $\tau$,
to the frequency range $\omega < 1\,$THz of the glycerol-water spectra,
see Fig.~\ref{fig:glyc_spectra}.

\begin{figure}
            \begin{center}
            \includegraphics[width=\columnwidth]{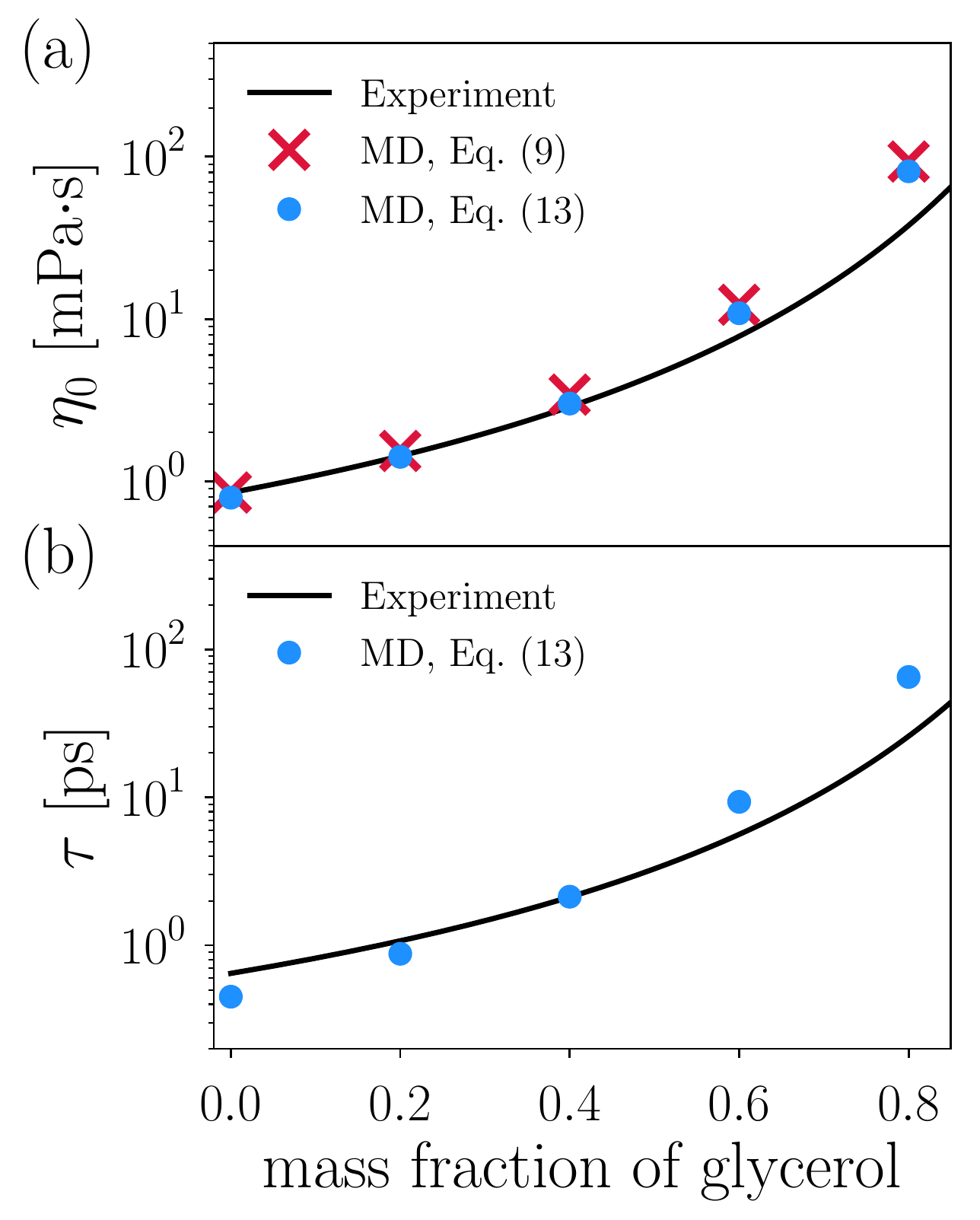}
            \end{center}
            \caption{\textbf{Comparison of experimental and simulation data for the viscoelasticity of glycerol solutions.}
            MD data points are obtained using the Green-Kubo formula Eq.~\eqref{eq:gk_all} (red crosses), or fitting
            a Maxwell model Eq.~\eqref{eq:maxwell}  to the full viscosity spectra (blue dots), 
            c.f.~Fig.~\ref{fig:glyc_spectra}.
             Experimental data for $\eta_0$ are taken from Cheng\cite{2418347/66N25GMQ}, 
             timescales are calculated using 
             $\tau = {\eta_0}/{K}$,
             with $\eta_0$ taken from
             Cheng\cite{2418347/66N25GMQ} and $K$ taken from Slie et al. \cite{2418347/VR5ET3DV}
	}
            \label{fig:glyc_lambda}
        \end{figure}

In Fig.~\ref{fig:glyc_lambda}, we compare the fitted viscosities and relaxation
 times $\eta_0$, $\tau$ to experimental
results\cite{2418347/66N25GMQ,2418347/VR5ET3DV}.
As can be seen, the viscosity spectrum of the simulated glycerol/water mixtures reproduces
very well both the low-frequency shear viscosity $\eta_0$ and the timescale $\tau$
for the Maxwell model.
In subplot (a) we additionally include the zero-frequency
viscosity, calculated using the Green-Kubo formula Eq.~\eqref{eq:gk_all}.
The good agreement with our Maxwell model fits
 serves as a validation that our approach of using Eq.~\eqref{eq:visc_freq_all} in conjunction
with force field MD simulations yields physically meaningful results.

Taking a closer look at the pure TIP4P/2005 water spectrum, 
shown in Fig.~\ref{fig:glyc_spectra} (a), which is very similar to the
previously obtained viscosity spectrum using the TIP4P force field\cite{2486906/M8ANX9ZV},
it transpires that a single Maxwell model, which features a monotonically decreasing
 real part and one peak in the imaginary part,
is not able to model the full viscosity spectrum observed
in the MD simulation, which contains
 two peaks in the imaginary part and a non-monotonic real part.
A Maxwell model only describes 
this spectrum
 for frequencies $f \lesssim 1$ THz,
as was noted before\cite{osullivan_viscoelasticity_2019}.

\section{The  shear inertia model}
\label{sec:ViscoelasticModeling}

To model the observed shear viscosity spectrum of pure water, 
we consider a general stress-strain relation,
which links off-diagonal components of stress
and strain rate tensors at $\Vect{k}=0$ via\cite{2486906/5SQ8ZI6Z}
\begin{equation}
	\label{eq:ViscoelasticNetworksStressStrain}
	\tilde{\sigma}_{\alpha \beta}(\omega) = 2(-\imath\omega)\tilde{\eta}(\omega)  \tilde{\epsilon}_{\alpha \beta}(\omega)
	\qquad \alpha \neq \beta,
\end{equation}
c.f.~Eq.~\eqref{eq:ConvolutedStressStrain}.
This relation is analogous to electrical circuit theory, where one is interested in
 the total complex impedance $\tilde{Z}(\omega)$ of a circuit, 
 which links time-dependent voltage $U$ and
time-dependent current $I$ via\cite{2486906/RU2TVWEF}
	$\tilde{U}(\omega) = \tilde{Z}(\omega) \tilde{I}(\omega)$.
In analogy to electrical networks, 
we use Eq.~\eqref{eq:ViscoelasticNetworksStressStrain}
to model the total complex shear viscosity $\tilde{\eta}(\omega)$
of a viscoelastic network \cite{2486906/5SQ8ZI6Z}. 

The building blocks for electrical circuits are resistor, capacitor and inductor, and each
of them has a characteristic complex impedance $\tilde{Z}(\omega)$, shown
in the left column of Table \ref{tab:CircuitElements}.
In viscoelasticity, usually only viscoelastic analogues of resistor and capacitor,
but not of inductor,
are considered.
The viscoelastic analogue of the resistor is the dashpot (both resistor and dashpot
dissipate energy), while the viscoelastic analogue of the capacitor is the spring (both
 capacitor and spring can store energy).
\begin{table*}[ht]
\caption{Electrical and viscoelastic circuit elements}
\centering
\begin{tabular}{ccc||ccc}
\multicolumn{3}{c||}{Electrical circuit} & \multicolumn{3}{c}{Viscoelastic network}\\
\hline
\hline
\multicolumn{2}{c}{Building block} & $\tilde{Z}(\omega)$ & $\tilde{\eta}(\omega)$ & \multicolumn{2}{c}{Building block} \\
\hline
Resistor &\includegraphics[width=1.0cm,height=0.3cm]{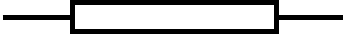} & $R$ & $\eta_0$ & \includegraphics[width=1.0cm,height=0.3cm]{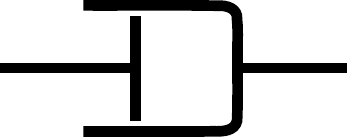}  & Dashpot\\
Capacitor &\includegraphics[width=1.0cm,height=0.3cm]{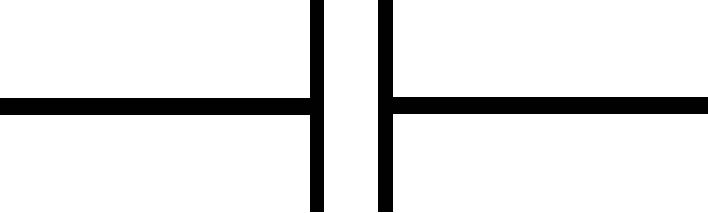} & $(-\imath \omega C)^{-1}$ & $K(-\imath\omega)^{-1}$ & \includegraphics[width=1.0cm,height=0.3cm]{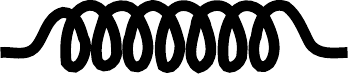}& Spring\\
Inductor &\includegraphics[width=1.0cm,height=0.3cm]{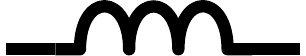} & $-\imath\omega L$ & $-\imath\omega L$& \includegraphics[width=0.7cm,height=0.4cm]{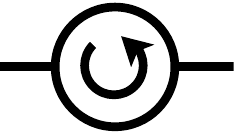}& Shear inertia \\
\end{tabular}
\label{tab:CircuitElements}
\end{table*}
The rules for calculating the total viscosity of a viscoelastic network, built up by serial and
parallel combination of viscoelastic building blocks, are illustrated in Table \ref{tab:ViscoelasticCombinationRules}\cite{2486906/5SQ8ZI6Z}\textsuperscript{,}\footnote{Note that these rules are exactly opposite of those for electrical circuits: For electrical circuits, the formula given here for parallel combination is used when one wants to calculate the total complex impedance for a serial combination of electrical circuit elements, and vice versa.}.

\begin{table*}[ht]
\caption{Combination rules for viscoelastic networks\cite{2486906/5SQ8ZI6Z}}
\centering
\begin{tabular}{cc|c|c}
\multicolumn{2}{c}{Combination} & Stresses \& Strains & Formula for $\tilde{\eta}_\mathrm{tot}(\omega)$\\
\hline
\hline
\pbox{2cm}{\vspace{0.1cm}Parallel}  & ~~\pbox{3cm}{\vspace{0.15cm}\includegraphics[width=1.5cm]{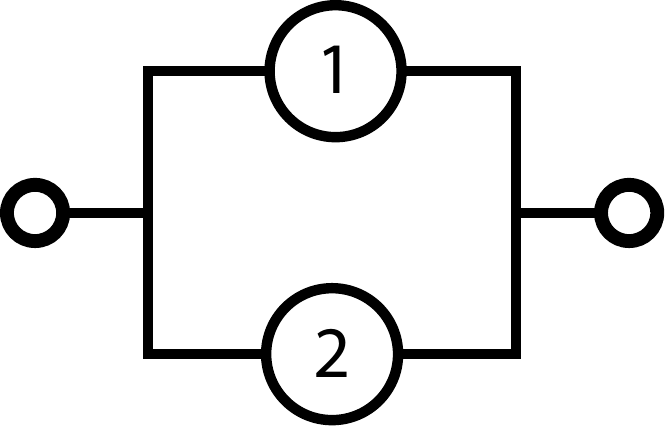}} ~~& \pbox{2.5cm}{$\sigma_{\mathrm{tot}} = \sigma_{1} + \sigma_{2}$\\ $\epsilon_{\mathrm{tot}} = \epsilon_{1} = \epsilon_{2}$} & $\tilde{\eta}_\mathrm{tot}(\omega) = \tilde{\eta}_1(\omega) + \tilde{\eta}_2(\omega)$\\[0.5cm]
\hline
& & &  \\
Serial &
 \pbox{3cm}{\vspace{0.02cm}\includegraphics[width=1.5cm,height=0.3cm]{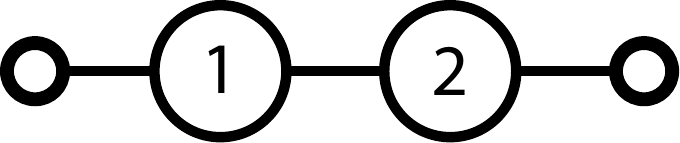} }
 & \pbox{2.5cm}{$\sigma_{\mathrm{tot}} = \sigma_{1} = \sigma_{2}$\\ $\epsilon_{\mathrm{tot}} = \epsilon_{1} + \epsilon_{2}$} & ~~$\tilde{\eta}_\mathrm{tot}(\omega) = \left(\tilde{\eta}^{-1}_1(\omega) + \tilde{\eta}^{-1}_2(\omega)\right)^{-1}$\\
\end{tabular}
\label{tab:ViscoelasticCombinationRules}
\end{table*}

\begin{figure}[ht]
  \centering
  \includegraphics[width=0.9\columnwidth]{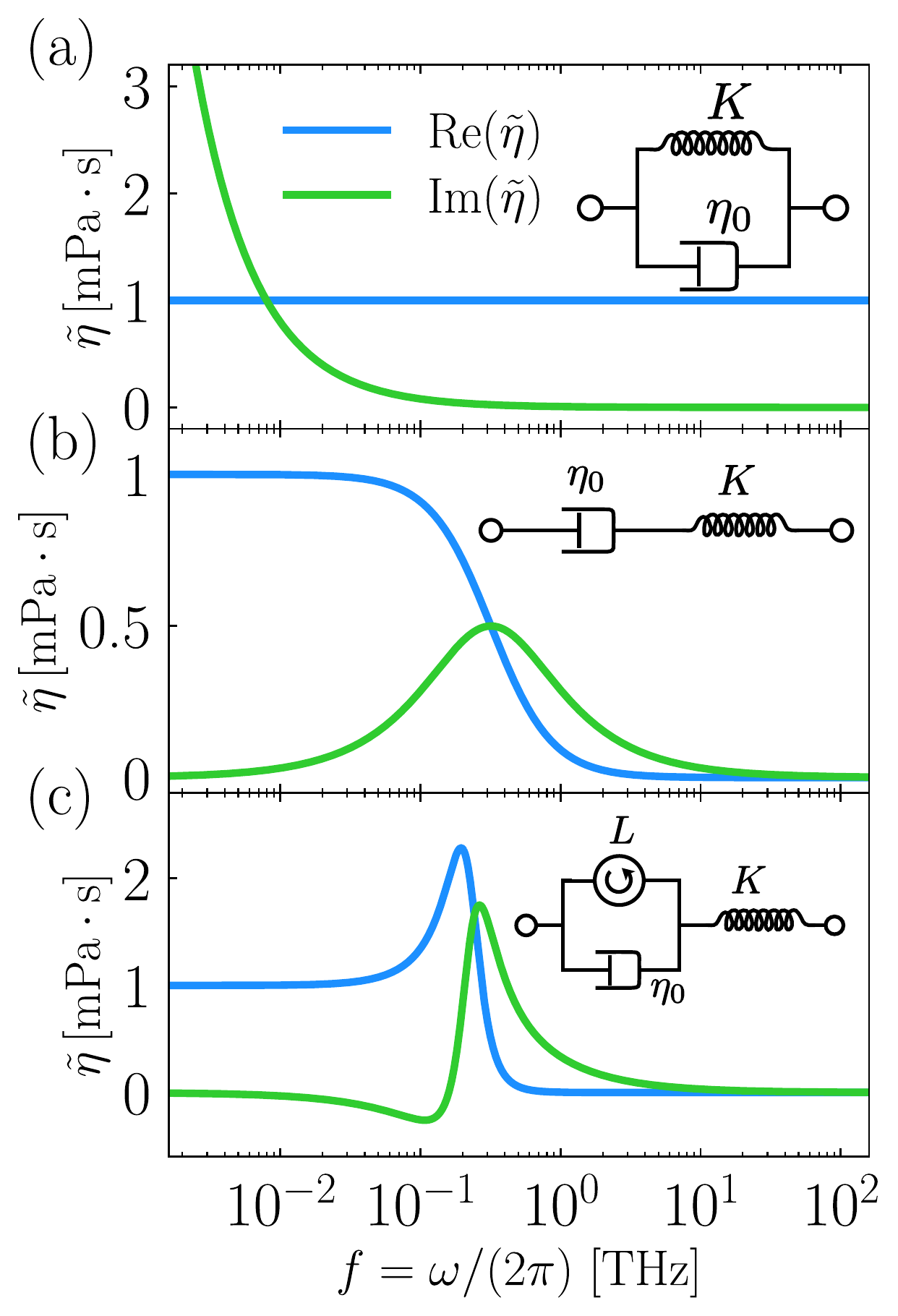}
    \caption{\textbf{Complex viscosities of viscoelastic materials.}
  \textbf{(a)} Complex viscosity of a Kelvin-Voigt model, 
 $ \tilde{\eta}(\omega) = \eta_0 + {K}/({-\imath\omega})$, 
  with $\eta_0 = 1\,\mathrm{mPa \cdot s}$, $K = 5 \cdot 10^{10}\,\mathrm{mPa}$.
\textbf{(b)} Complex viscosity of a Maxwell model, Eq.~\eqref{eq:maxwell}, with $\eta_0 = 1\,\mathrm{mPa \cdot s}$, $K = 2 \cdot 10^{12}\,\mathrm{mPa}$.
 \textbf{(c)} Complex viscosity of a shear inertia model, 
   Eq.~\eqref{eq:ViscoelasticRLCResponse}, with $\eta_0 = 1~\mathrm{mPa\cdot s}$, $K = 2\cdot 10^{12}~\mathrm{mPa \cdot s}$, $L = 10^{-12}~\mathrm{mPa \cdot s^2}$.
}
     \label{fig:StandardMaterials}
\end{figure}

In Fig.~\ref{fig:StandardMaterials} (a), (b), we illustrate the viscoelastic response of  
two standard viscoelastic materials.
The Kelvin-Voigt material, shown in Fig.~\ref{fig:StandardMaterials} (a), 
is a parallel combination of a dashpot (viscosity $\eta_0$) and a spring (elastic modulus $K$);
it has a complex viscosity
	$\tilde{\eta}(\omega) = \eta_0 + {K}/({-\imath\omega})$,
and models a viscoelastic solid.
The Maxwell material,
defined in Eq.~\eqref{eq:maxwell} and shown in 
Fig.~\ref{fig:StandardMaterials} (b),
 is
 a serial combination of a dashpot and a spring, and models 
 a viscoelastic fluid;
the dashpot viscosity $\eta_0$ and spring elastic modulus $K$ are related to the
relaxation timescale $\tau$ via 
$\tau = {\eta_0}/{K}$.

As can be seen from Fig.~\ref{fig:StandardMaterials} (a), (b), 
both Kelvin-Voigt and Maxwell model do not exhibit a 
 real part of the complex viscosity
 that is increasing as a function of frequency.
To model the pure water spectrum, and in particular the non-monotonic real part
of the viscosity in Fig.~\ref{fig:glyc_spectra} (a), 
we introduce the viscoelastic analogue of the inductor, which we
 denote by a circle containing a curved arrow, see right column of Table \ref{tab:CircuitElements}.
 This new circuit element models
  stress generated by shear acceleration.
Indeed, substituting $\LT{\eta}(\omega) = - \imath \omega L$
  into Eq.~\eqref{eq:ViscoelasticNetworksStressStrain} and using Eq.~\eqref{eq:RateOtStrainTensor}, it follows that
  \begin{equation}
  \label{eq:shear_inertia}
\sigma_{\alpha\beta} = 2 L    \ddot{\epsilon}_{\alpha \beta} = 
L   \left( \frac{\partial \dot{v}_\alpha}{\partial x_\beta} + \frac{\partial \dot{v}_\beta}{\partial x_\alpha}\right),
  \end{equation}
meaning that the stress is proportional to the shear acceleration. 
Substituting Eq.~\eqref{eq:shear_inertia} into the 
momentum conservation Eq.~\eqref{eq:LinearizedMomentumConservation},
the resulting force can be interpreted as contributing an effective induced mass due to velocity gradients
in the fluid.
A possible microscopic origin of such an effect 
 is the coupling between shear and rotational degrees 
of freedom of individual fluid particles (vorticity-spin coupling\cite{bonthuis_electrokinetics_2010}).

Using the shear inertia circuit element, 
we
 consider the viscoelastic network 
shown in Fig.~\ref{fig:StandardMaterials} (c),
which we call the shear inertia model.
As we show now, this is a generalization of the Maxwell model which includes shear inertial effects.
Using the viscosities of the individual network elements, c.f.~Table \ref{tab:CircuitElements},
and the combination rules, c.f.~Table \ref{tab:ViscoelasticCombinationRules}, the total viscosity
of the shear inertia model in Fig.~\ref{fig:StandardMaterials} (c) is found to be
\begin{equation}
	\label{eq:ViscoelasticRLCResponse}
 \tilde{\eta}(\omega) = \eta_0 \frac{1 + (-\imath\omega)\tau_{m}}{1 + (-\imath\omega)\tau_{0}^2/\tau_m + (-\imath\omega)^2 \tau_{0}^2},
\end{equation}
where
	$\tau_{m} = {L}/{\eta_0}$, 
	$\tau_0^2 = {{L}/{K}}$.
From Eq.~\eqref{eq:ViscoelasticRLCResponse} it can be seen that this is an extension of the Maxwell model, Eq.~\eqref{eq:maxwell}, which is recovered in the limit $\omega \tau_m \rightarrow 0$,
$\omega \tau_0 \rightarrow 0$,
at finite $\omega \tau_0^2/\tau_m \equiv \omega \tau$.

An important difference between the Maxwell and the
shear inertia model is that the latter features a non-montonic
real part in the complex viscosity;
from equating its derivative with zero,  it follows
 that the real part of $\LT{\eta}(\omega)$
 defined by Eq.~\eqref{eq:ViscoelasticRLCResponse} 
  is non-monotonic 
if and only if 
\begin{equation}
	\label{eq:ViscoelasticRLCConditionForMaximum}
	 \left(\frac{\tau_0}{\tau_m}\right)^2 = \frac{\eta_0^2}{KL} < 1.
\end{equation}
In Fig.~\ref{fig:StandardMaterials} (c), we show the complex viscosity 
of the \InertialMW for parameters $\eta_0 = 1~\mathrm{mPa\cdot s}$, 
$K = 2\cdot 10^{12}~\mathrm{mPa\cdot s}$, $L = 10^{-12}~\mathrm{mPa \cdot s^2}$.
In the low-frequency limit the viscosity spectrum approaches that of a Maxwell model, 
with constant real part and negligible imaginary part.
Also, in the high frequency limit, both real and imaginary parts vanish, as they do in the Maxwell model.
However, different from the Maxwell model, the real part of the viscosity is non-monotonic and
has a maximum, while the imaginary part is negative for small frequencies.

\section{A viscoelastic model for pure water}

\begin{figure*}[ht]
        \centering
        \includegraphics[width=\textwidth]{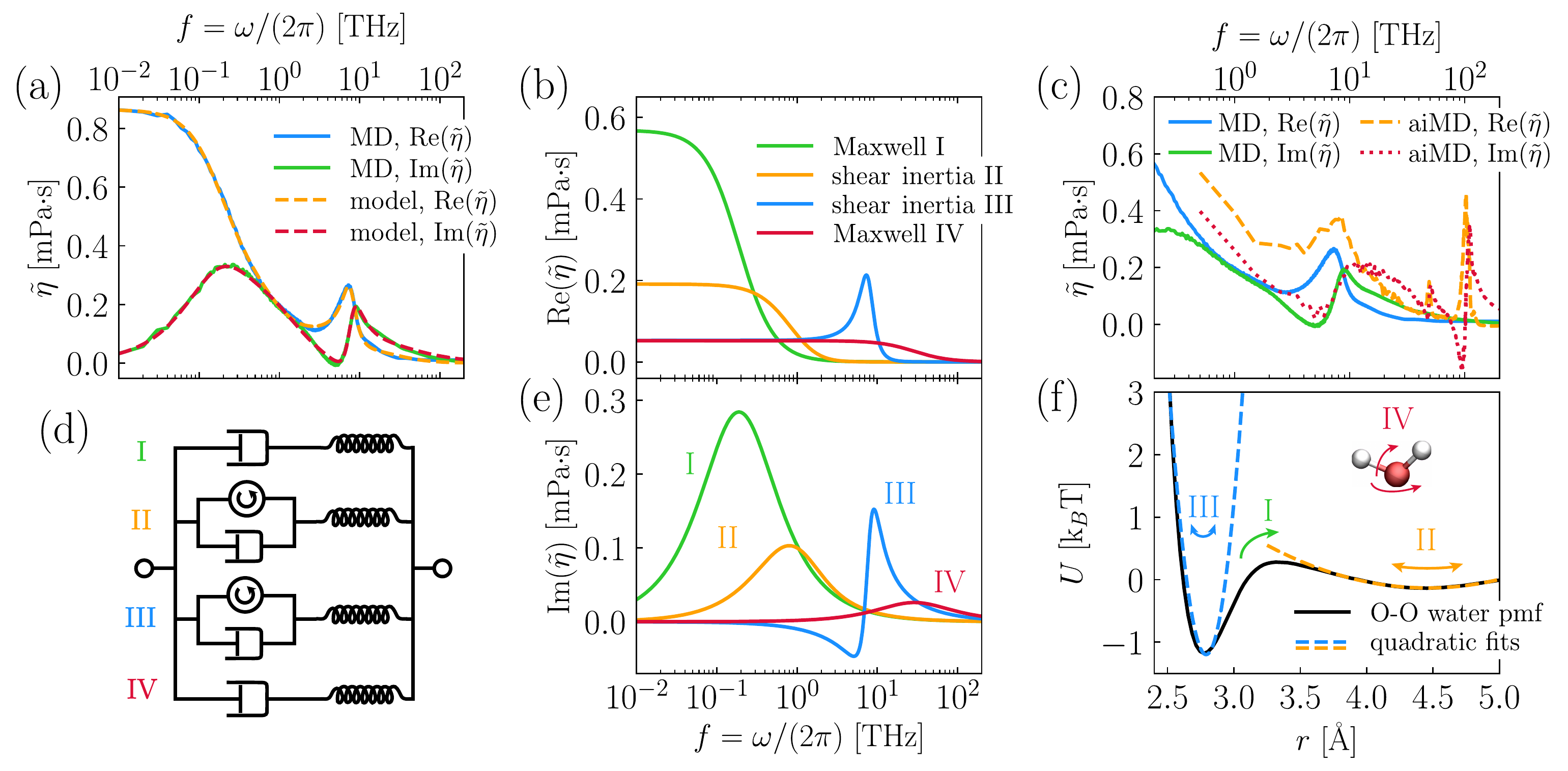}
                \caption{\textbf{Viscosity spectrum of pure water.}
                \textbf{(a):} 
                		Complex viscosity $\tilde{\eta}$ from MD simulations of TIP4P/2005 water,
		 	calculated using Eq. (\ref{eq:visc_freq_all}),
			together with a fit of the viscoelastic network shown in subplot (d)
			 with complex viscosity Eq.~\eqref{eq:ThreeMaxwellLC}.
   			The fitting parameters are given in Table \ref{tab:ModelFit}.
	    \textbf{(b), (e):}
	    		 Real and imaginary parts of the individual constituents of the fitted viscosity from subplot (a);
	\textbf{(c):}
			Complex viscosity $\tilde{\eta}$ from MD simulations of TIP4P/2005 water (replot of subplot (a)),
			together with the viscosity calculated from ab-initio molecular dynamics (aiMD) simulations using 
			Eq. (\ref{eq:ViscosityApproximateFormula2}) at finite $k =  4.016$ nm${}^{-1}$.
	  \textbf{(d):} 
	  		Viscoelastic network used for modeling the shear response of pure water.
               		The viscoelastic network is a parallel combination of two shear inertia models and
			 two Maxwell models, 
                		the resulting complex viscosity is given by Eq.~\eqref{eq:ThreeMaxwellLC};
	\textbf{(f):} 
			Black line: Potential of mean force $U(r)$ for the oxygen-oxygen distance in TIP4P/2005 water.
            		Blue and yellow lines: Harmonic potentials, fitted around 
		the first and second minima of the pmf.
	            	The colored arrows indicate the microscopic interpretation of the constituents of the
			viscoelastic model shown in subplot (d), namely escape from the nearest neighbor pmf minimum (I), 
			oscillations around the minima (II, III) and librations of individual water molecules (IV). 	
                }
                \label{fig:SpectraTogether}
\end{figure*}

To model the pure water spectrum, we employ a parallel combination of two Maxwell models 
and two shear inertia models, illustrated in Fig.~\ref{fig:SpectraTogether} (d).
The total complex viscosity of this network is given by
\begin{align}
	\tilde{\eta}(\omega) &= \sum_{j=\mathrm{II,III}} \eta_{0,j} \frac{1 -\imath \omega\tau_{m,j}}{1 -\imath\omega\tau_{0,j}^2/\tau_{m,j} -\omega^2 \tau_{0,j}^2} + \nonumber \\
	\label{eq:ThreeMaxwellLC}
	&\qquad +  \sum_{j=\mathrm{I, IV}} \frac{\eta_{0,j}}{1 - \imath \omega \tau_{j}}.
\end{align}
The result of a fit of this model to the MD data is shown in Fig.~\ref{fig:SpectraTogether} (a),
 the resulting parameters are given in Table \ref{tab:ModelFit}.
As Fig.~\ref{fig:SpectraTogether} (a) demonstrates, this viscoelastic network
is able to reproduce the MD spectrum very well.

In the SM\cite{supplement} we show that fitting three Maxwell models and one shear inertia model 
yields results of similar quality. 
Our choice of two Maxwell models and two shear inertia models
is mainly motivated by the microscopic interpretation of the four relaxation modes 
by which a given strain rate creates stresses, as we will explain in the next section.

\begin{table}[ht]
\caption{Parameters for the viscoelastic model Eq.~(\ref{eq:ThreeMaxwellLC}), 
resulting from a fit to results for TIP4P/2005 water,
 c.f.~Fig.~\ref{fig:SpectraTogether} (a-c), (e).
Timescales are 
  converted to frequencies via $f = (2 \pi \tau)^{-1}$ for ease of comparison with Fig.~\ref{fig:SpectraTogether}.}
\centering
\begin{tabular}{c|rl|l}
Parameter &  Value &  & Interpretation\\
\hline
\hline
$\eta_{0,\mathrm{I}}$ &  0.568 & {mPa}$\cdot$s ~& HB network \\
$(2\pi\cdot \tau_{\mathrm{I}})^{-1}$ & 0.19 & $\mathrm{THz}$    & topology changes\\
\hline
$\eta_{0,\mathrm{II}}$ &   0.019 & {mPa}$\cdot$s & O-O-O vibrations\\
$(2\pi\cdot \tau_{m,\mathrm{II}})^{-1}$ & 2.68 & $\mathrm{THz}$ & (HB bending)\\
$(2\pi\cdot \tau_{0,\mathrm{II}})^{-1}$ & 1.27 & $\mathrm{THz}$ & \\
\hline
$\eta_{0,\mathrm{III}}$ &   0.053 & {mPa}$\cdot$s & O-O vibrations\\
$(2\pi\cdot \tau_{m,\mathrm{III}})^{-1}$ & 4.03 & $\mathrm{THz}$ & (HB stretching)\\
$(2\pi\cdot \tau_{0,\mathrm{III}})^{-1}$ & 7.83 & $\mathrm{THz}$ & \\
\hline
$\eta_{0,\mathrm{IV}}$  &  0.052 & {mPa}$\cdot$s & Librations\\
$(2\pi\cdot \tau_{\mathrm{IV}})^{-1}$ &  28.92 & $\mathrm{THz}$ &  \\
\end{tabular}
\label{tab:ModelFit}
\end{table}

\section{Microscopic interpretation of the pure water viscosity spectrum}
\label{sec:MicroscopicOrigin}

We now give a microscopic interpretation for each of 
the constituents of the fit of Eq.~\eqref{eq:ThreeMaxwellLC}
to the MD viscosities, c.f.~Fig.~\ref{fig:SpectraTogether} (a), (b), (e).

We first calculate the radial distribution function $g(r)$ for oxygen atoms and use it to obtain
 the potential of mean force (pmf) $U(r)$ for the effective oxygen-oxygen interaction via Boltzmann inversion,
\begin{equation}
	\label{eq:PmfFormula}
	U(r) = -k_{\mathrm{B}}T \,\ln\left[g(r)\right],
\end{equation}
see Fig.~\ref{fig:SpectraTogether} (f).
The pmf shows a primary 
minimum at an oxygen-oxygen distance of approximately $r \approx 2.8$ \AA, corresponding to
 two hydrogen-bound nearest neighbor water molecules. 

We start with the microscopic interpretation of Maxwell model I, which shows a peak in
the elastic response at $f\approx 0.2\,\mathrm{THz}$, c.f.~Fig.~\ref{fig:SpectraTogether} (e).
This elastic response has been observed experimentally for water\cite{2418347/8BRIQ6QX}, and in the context
of a Yukawa liquid it was recently uncovered that this peak is linked to 
nearest-neighbor escape barrier hopping within
the fluid\cite{2486906/MUDD6K49,2486906/XC5WZN3B}, with the peak frequency 
corresponding to the 
inverse lifetime of the nearest-neighbor topology of a given molecule.
Mean first-passage times for the dissociation of a water-water pair from the nearest neighbor pmf minimum for SPC/E water are about\cite{2486906/7KPB6RN6,2486906/QJ5VJU9Q}
 $\tau_{\mathrm{escape}} \approx 4\,\text{--}\,5\,\mathrm{ps}$, which translates to frequencies
 $1/\tau_{\mathrm{escape}}\approx 0.20\,\text{--}\,0.25\,\mathrm{THz}$,
in very good 
agreement with the position of the leftmost peak of the imaginary part in Fig.~\ref{fig:SpectraTogether} (a), (c).
A comparison of these SPC/E results with our TIP4P/2005 data is legitimate, because, as
we show in the SM\cite{supplement}, the viscosity spectrum of SPC/E water is very similar to
 the TIP4P/2005 spectrum shown in Fig.~\ref{fig:SpectraTogether}.
In particular the lowest-frequency peak is located at a comparable frequency.

The second feature we interpret microscopically is the shear inertia model III. 
In Table \ref{tab:Frequencies} we give frequencies for various modes in liquid water obtained from experiments and simulations.
Both Raman- and infrared (IR) spectroscopy find hydrogen bond stretch vibrations 
at frequencies $f \approx 4.5\,\text{--}\,5.5$ THz, 
while simulation works find them slightly higher, 
at 6 THz (aiMD\cite{2486906/35XBWCHB}) and 6.9 THz (TIP4P/2005f\cite{2486906/DGPPXZVD}), respectively.
This suggests that the microscopic origin of the corresponding resonance in our spectrum are
 vibrations of hydrogen-bound water pairs around the minimum at $r \approx 2.8 \,\mathrm{\AA}$ in Fig.~\ref{fig:SpectraTogether} (f).
 Multiplying Eq.~\eqref{eq:ViscoelasticRLCResponse} by the denominator of the right-hand side,
 and performing an inverse Fourier transform, 
it can be seen that $\eta(t)$ is the solution of the damped harmonic oscillator equation
\begin{equation}
	\label{eq:InertialMaxwellModelDGL}
	\ddot{\eta}(t) + \frac{1}{\tau_m} \dot{\eta}(t) + \frac{1}{\tau_0^2} \,\eta(t) =
	0
\end{equation}
with initial conditions $\eta(0) = \eta_0\tau_m/\tau_0^2$, $\dot{\eta}(t) = 0$.
With the fitted parameters for shear inertia model III, given in Table \ref{tab:ModelFit}, it follows 
that the damped harmonic oscillator solution to Eq.~\eqref{eq:InertialMaxwellModelDGL}  
is indeed
an underdamped oscillation,
which is why an extension of the standard (overdamped)
Maxwell model is required to describe this feature.
The resonance frequency of 
the underdamped harmonic oscillator defined by Eq.~\eqref{eq:InertialMaxwellModelDGL}
evaluates to
\begin{equation}
\label{eq:fr_shear_inertia_III}
	f_r = \frac{1}{2\pi} \sqrt{ \frac{1}{\tau_0^2} - \frac{1}{4\tau_m^2}} \approx 6.7~\mathrm{THz},
\end{equation}
which is close to the values for hydrogen-bond stretching vibrations found in the literature,
see Table \ref{tab:Frequencies}.
Furthermore, the timescales $\tau_0$, $\tau_m$, obtained from the fitted shear inertia model III
are in agreement with
the intuitive picture of a hydrogen-bound water pair oscillating around the first minimum of the pmf
shown in Fig.~\ref{fig:SpectraTogether} (f),
as we will explain next.
To proceed, according to Eq.~\eqref{eq:InertialMaxwellModelDGL}, the frequency
for undamped oscillations obtained from the shear inertia model III fit
is given by
\begin{equation}
	\label{eq:iM3FreqFit}
	f_{\mathrm{osc,III}} = \frac{1}{2 \pi \tau_{0,\mathrm{III}}}
					 \approx 7.83~\mathrm{THz},
\end{equation}
and not very different from the result including damping in Eq.~\eqref{eq:fr_shear_inertia_III}.
The frequency Eq.~\eqref{eq:iM3FreqFit} is comparable to the 
 frequency for an undamped harmonic
 oscillation of a bound water pair around the first minimum of the pmf,
\begin{equation}
	\label{eq:FrequencyHO}
	f_{\mathrm{ho}} = \frac{1}{2 \pi}\sqrt{\frac{k}{m}} \approx 8.82~\mathrm{THz},
\end{equation}
where the force constant $k = 110.85\,k_{\mathrm{B}}T/\mathrm{\AA}^2$ is obtained by fitting a quadratic potential
\begin{equation}
	U(r) = U(r_0) + \frac{k}{2} (r-r_0)^2 
\end{equation}
to the minimum at $r_0 \approx 2.8$\,\AA, c.f.~the blue dashed curve in Fig.~\ref{fig:SpectraTogether} (f), and 
for $m$ we use the reduced water mass $m = m_{\mathrm{water}}/2 = 9\,$amu,
appropriate for relative oscillations of two rigid water molecules.
From the damping term in Eq.~\eqref{eq:InertialMaxwellModelDGL},
the effective friction coefficient $\gamma$  for 
 a vibrating water molecule pair
within the hydrogen bond network
is estimated as
 \begin{equation}
 \label{eq:friction_shear_inertia_III}
 	\gamma \sim m /\tau_m \approx 2 \pi \cdot 9 \, \mathrm{amu} \cdot 4.03\,\mathrm{THz} \approx 
	0.38\, \frac{\mathrm{pN\,ns}}{\mathrm{nm}}.
\end{equation}
This value is almost one order of magnitude smaller than the friction coefficient of a 
diffusing water molecule, $\gamma \approx  1.62\, \mathrm{pN\,ns/nm}$\cite{2486906/7KPB6RN6},
which physically makes sense because a diffusing water molecule  is expected to experience more resistance
to motion
as compared to a particle oscillating within a local potential minimum.
Relative oscillations of hydrogen-bound water molecules
contribute to the shear viscosity because of the polarity of an individual water molecule,
which couples translation and rotation of individual molecules within the hydrogen-bond network;
indeed, as we show in the SM\cite{supplement}, 
upon turning off the electrostatic interactions between the 
water molecules,
the feature in the viscosity spectrum disappears.
Note furthermore that our interpretation is consistent with the fact that the peak at about $9~\mathrm{THz}$
 disappears as the glycerol mass fraction is increased, c.f.~Fig.~\ref{fig:glyc_spectra}, because diffusing glycerol molecules
hinder the hydrogen bond network of water\cite{2486906/Z8B85W6N}.
 We thus conclude that the microscopic origin of shear inertia model III are hydrogen bond vibrations of neighboring water molecules.

\begin{table*}[ht]
\caption{Frequencies for various resonances of liquid water.
All frequencies are given in THz. }
\centering
\begin{tabular}{p{2.5cm}||p{2.2cm} | p{3.2cm} | p{1.5cm} | p{2.5cm} | p{3.7cm}}
     & Raman spectroscopy\cite{2486906/MD356JZF}
     & Infrared\newline spectroscopy (IR)\cite{2486906/QU86HKME,2486906/ZW2ASP7U}
     & IR from aiMD\cite{2486906/35XBWCHB}
     & IR from MD (TIP4P/2005f)\cite{2486906/DGPPXZVD}
     & viscosity from MD (TIP4P/2005, this work) \\
\hline
\hline
 O-O vibrations \newline
 (HB stretching) &
 4.7 &
 $\approx$ 5.1-5.5 &
 6.0 &
6.9 &
6.7 \\
\hline
 O-O-O vibrations \newline
 (HB bending) &
 2.0 &
 1.5 &
 2.4 &
 1.5 &
 1.7\\
\hline
 Librations &
$\approx$ 12-24 &
$\approx$ 12-21 &
$\approx$ 18-24 &
$\approx$ 17 
& 28.9\\
\end{tabular}
\label{tab:Frequencies}
\end{table*}

Similarly, we interpret shear inertia model II as oscillations around the second
minimum of the pmf shown in Fig.~\ref{fig:SpectraTogether} (f), at around $r_0 \approx 4.5\,$\AA, corresponding to
 oxygen-oxygen-oxygen (O-O-O) vibrations within the hydrogen bond
network of water.
Note that, in contrast to shear inertia model III, 
now the fitted damped harmonic oscillator Eq.~\eqref{eq:InertialMaxwellModelDGL} 
is overdamped,
so that O-O-O hydrogen bond vibrations are actually overdamped.
As we show in the SM\cite{supplement}, using a standard Maxwell model for feature II is also
 possible.
Indeed,
approximating the overdamped mode Eq.~\eqref{eq:ViscoelasticRLCResponse}
by a Maxwell model with relaxation time scale $\tau = \tau_0^2/\tau_m$, we find a
 resonance frequency $f_r \approx (2 \pi \tau)^{-1} \approx 1.7$ THz, which is
 close to the literature values for hydrogen-bond bending
  vibrations, c.f.~Table \ref{tab:Frequencies}.
Both experiments and simulations locate these 
vibrations at frequencies slightly higher than our MD water model\cite{2486906/MD356JZF,2486906/QU86HKME,2486906/ZW2ASP7U,2486906/DGPPXZVD}, see Table \ref{tab:Frequencies}.
While the fitted shear inertia model II  describes
an overdamped oscillator,
employing a shear inertia model instead of a Maxwell model
 allows us to estimate the effective friction coefficient for 
O-O-O vibrations,
as
 \begin{equation}
 	\gamma \sim m /\tau_m \approx 2 \pi \cdot 9 \, \mathrm{amu} \cdot 2.68\,\mathrm{THz} \approx 
	0.25\, \frac{\mathrm{pN\,ns}}{\mathrm{nm}},
\end{equation}
 where we again use the reduced mass $m = m_{\mathrm{water}}/2 = 9\,\mathrm{amu}$ 
 as an estimate for the inertia, 
 based on the picture that during O-O-O bending vibrations, the middle molecule does not move significantly.
This value for the molecular friction $\gamma$
 is of the same order of magnitude as the one obtained for shear inertia model III,
c.f.~Eq.~\eqref{eq:friction_shear_inertia_III},
and is also considerably smaller than the friction coefficient of a 
diffusing water molecule, $\gamma \approx  1.62\, \mathrm{pN\,ns/nm}$\cite{2486906/7KPB6RN6}.
Note finally that also for shear inertia model II, 
the frequencies obtained from the fit are in agreement with 
expectations from the pmf in Fig.~\ref{fig:SpectraTogether} (f).
Indeed, a quadratic fit to the second minimum of the pmf, shown as the yellow dashed 
line in the figure, leads to an undamped oscillation frequency
\begin{equation}
	\label{eq:FrequencyHO2}
	f_{\mathrm{ho}} = \frac{1}{2 \pi}\sqrt{\frac{k}{m}} \approx 0.81~\mathrm{THz},
\end{equation}
 where as above we use the reduced mass $m = m_{\mathrm{water}}/2$ as an estimate for the inertia. 
The fitted shear inertia model II yields the comparable
frequency
\begin{equation}
	  \label{eq:iM2FreqFit}
	f_{\mathrm{osc,II}} = \frac{1}{2 \pi \tau_{0,\mathrm{II}}}
					 \approx 1.27~\mathrm{THz}.
\end{equation}

The highest frequency features, described by Maxwell model IV, we interpret as
librational excitations, i.e.~rotational vibrations of individual water molecules within the
force field of their surrounding molecules.
To show this, we in the SM consider orientational spectra calculated for individual water molecules,
which show a peak at frequencies comparable to
 the elastic response frequency of Maxwell model IV\cite{supplement}.
Note that spectroscopy locates such vibrations at frequencies
 considerably higher than O-O vibrations, c.f.~Table \ref{tab:Frequencies}.
That inertial effects can be neglected here is consistent with the fact that for the rotational motion of SPC/E water,
 inertia starts to dominate only at much higher frequencies, namely at around 90 THz\cite{YannThesis}.

In addition to our results obtained from TIP4P/2005, a classical rigid water model, 
 we calculate the viscosity spectrum for pure water also from ab-initio MD (aiMD) 
simulations,
see SM for details\cite{supplement}.
Since no stress tensor is available for the aiMD data, we use Eq.~(\ref{eq:ViscosityApproximateFormula2})
 at finite $k =  4.016$ nm${}^{-1} = 2\pi/L$, 
where $L=1.56$ nm is the edge length of the cubic simulation box.
 A comparison of aiMD and TIP4P/2005 spectra
 (the latter as before at $k=0$) is shown in Fig.~\ref{fig:SpectraTogether} (c).
Qualitative deviations between the spectra appear at frequencies above 50 THz, 
where intramolecular degrees of freedom\cite{2486906/MD356JZF,2486906/35XBWCHB}
(OH stretching, OH bending), which are not accounted for in a rigid water model,
become relevant.
Up to these frequencies, however, the spectra show the same features, in particular the peak at around 9\,THz with its high-frequency shoulder is present in both the aiMD and the force field MD spectra;
this confirms that the spectrum obtained from force field MD is
indeed accurate up to about 50\,THz.
This finding is in agreement with the recent comparison of aiMD and force field MD results for the linear
absorption of pure water from 1\,MHz to 100\,THz \cite{carlson_exploring_2020}.

\section{Conclusions}

In this work we calculate the viscoelastic properties of both pure water and water-glycerol mixtures
from force field MD simulations. 
For water-glycerol mixtures, we find very good agreement
of the viscosity and the relaxation time with experimental data \cite{2418347/VR5ET3DV,2418347/66N25GMQ}, 
for pure water our spectrum
agrees with previous theoretical results\cite{2486906/M8ANX9ZV} 
and a spectrum calculated from aiMD simulations at finite wave numbers.
Using an extension of the Maxwell model for viscoelastic fluids, which includes shear inertial effects,
we propose a functional form for the shear viscosity of pure water that is able to 
describe the force field MD spectrum over the entire frequency range considered.
By comparing to Raman and IR spectra, as well as to other observables calculated from our MD data,
we subsequently identify the molecular processes underlying this spectrum as water network topology
changes, collective vibrations of three water molecules,
hydrogen-bond stretch vibrations of water pairs,
  and librational excitations of individual water molecules.

The viscoelastic circuit we propose describes the short-time non-Markovian behavior of water
in the picosecond regime,
where the standard Newtonian fluid stress-strain relation
becomes a poor approximation to the actual dynamics of the fluid.
Only the viscoelastic response at the longest time scales
  discussed here has been measured until now\cite{2418347/VR5ET3DV,2418347/8BRIQ6QX},
so that our results present a challenge for future experimental investigation.

To obtain an even more detailed picture of the breakdown of the Newtonian fluid picture at small scales,
a possible next step would be to also systematically study short-distance non-local effects by considering 
viscosity kernels at finite wave vector $k$\cite{2486906/M8ANX9ZV,2486906/PFFPU8MS}.

Note finally that recently it was argued that phonons can travel along the hydrogen bond network,
and that elastic peaks in the range of tens of THz should be interpreted as such phonons\cite{2486906/8ZWTTQ5U},
which is consistent with our interpretation of these peaks as bond vibrations.
Since at time scales smaller
 than those of water network topology changes, one could think of water
as a static network, possibly with defects, starting from a tetrahedral
lattice and calculating the corresponding viscoelastic response\cite{2486906/7UNS7T4J} might yield further
insights into the high-frequency viscoelastic properties of water.

\begin{acknowledgments}
We acknowledge funding from the DFG via SFB 1114, Project C2.
\end{acknowledgments}

%

\end{document}